\documentclass[aps,prl,reprint,groupedaddress,showpacs]{revtex4-1}
\usepackage{graphicx}
\usepackage{dcolumn}
\usepackage{bm}
\usepackage{amssymb}
\usepackage{mathrsfs}
\usepackage{CJK}
\begin{document}

\title{Heterodyne detection enhanced by quantum correlation}

\author{Boya Xie}
\author{Sheng Feng}
\email[]{fengsf2a@hust.edu.cn}


\affiliation{Hubei Key Laboratory of Modern Manufacturing Quality Engineering, School of Mechanical Engineering, Hubei University of Technology, Wuhan 430068, China}

\date{\today}

\begin{abstract}
Heterodyne detectors as phase-insensitive (PI) devices have found important applications in precision measurements such as space-based gravitational-wave (GW) observation. However, the output signal of a PI heterodyne detector is supposed to suffer from signal-to-noise ratio (SNR) degradation due to image band vacuum and imperfect quantum efficiency. Here we show that the SNR degradation can be overcome when the image band vacuum is quantum correlated with the input signal. We calculate the noise figure of the detector and prove the feasibility of heterodyne detection with enhanced noise performance through quantum correlation. This work should be of great interest to ongoing space-borne GW signal searching experiments.
\end{abstract}

\maketitle

\noindent Heterodyne detection is a powerful tool to capture low-frequency weak signals ($\le$ 1 kHz) carried by optical beams, e.g., interferometric signals generated by gravitational-wave (GW) disturbances at space-borne observatories \cite{armano2016,luo2016,babak2017}.
Nevertheless, traditional heterodyne detectors as phase-insensitive (PI)  devices inevitably suffer from signal-to-noise ratio (SNR) degradation, caused by image band vacuum at its input \cite{shapiro1979,yuen1983,yamamoto1986,caves1994} and imperfect-quantum-efficiency induced vacuum at its output \cite{bencheikh1995}. As space-based GW detection systems are approaching their quantum noise limits \cite{Sesana2016}, the quantum noise property of heterodyne detectors will become an important limiting factor for further system sensitivity improvement in the near future.
Therefore, conquering the SNR degradation in heterodyne detection, if possible, may considerably benefit GW signal searching experiments because the volume of space that is probed for potential GW sources increases as the cube of the strain sensitivity.

According to the current heterodyne detection theory \cite{shapiro1979,yuen1983}, 3 dB noise penalty occurs in heterodyne detection due to the image band vacuum at the detector's input. It was suggested that the 3 dB heterodyne noise might be suppressed by injection of light in two-photon coherent states at the degenerate frequency of the image band vacuum into the detector \cite{yuen1980}, which unfortunately has never been implemented so far. Using amplitude-squeezed local oscillator, the quantum noise of a one-port heterodyne detector may be reduced \cite{li1999}, yet the problem of the 3 dB extra heterodyne noise due to the image band vacuum was not addressed. A phase-sensitive (PS) heterodyne detector with a bichromatic local oscillator has proven to be noiseless \cite{marino2007,liw2015,feng2016,liuf2017,xie2018b}, but its phase sensitivity \cite{feng2016} requires phase control for the input signal that is intractable in the detection scheme of ongoing space-based GW experiments where violent disturbance to the phase of the input signal is inevitable \cite{armano2016,luo2016,babak2017}.

\begin{figure}
\includegraphics[width=7cm]{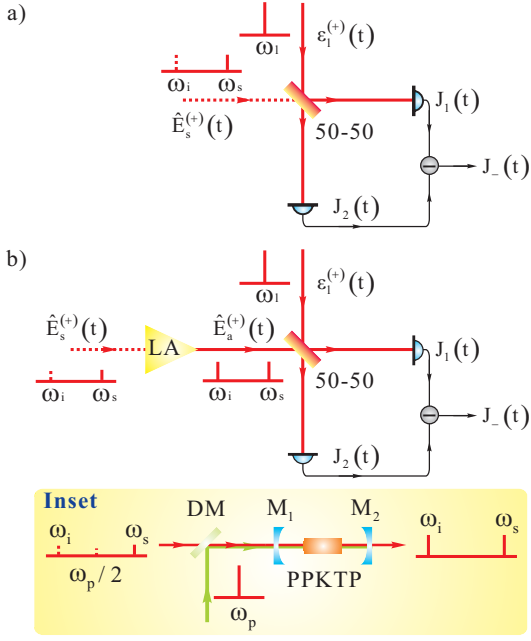}
\caption{(color online) Schematics for heterodyne detection of light. (a) The signal mode enters into the detector together with an unexcited mode (image band vacuum) that gives rise to 3 dB heterodyne noise \cite{shapiro1979,yuen1983,caves1994}. (b) Before being received by the detector, the signal light passes through a parametric amplifier where quantum correlation \cite{kong2013} is generated between the signal mode and the image band vacuum for reduction of the 3 dB heterodyne noise. $\omega_{\rm s,i,l}$: Angular frequency of the signal/image band/ local oscillator mode. $\hat{E}_{\rm s,a}^{(+)}(t)$: Quantum field of signal/amplified light beam. $\varepsilon_{\rm l}^{(+)}(t)$: Classical field of local oscillator light beam. 50-50: Balanced beamsplitter. LA: Linear amplifier. $J_-(t)\equiv J_2(t)-J_1(t)$: Average differential photocurrent signal from the detector. Inset: a typical parametric amplifier consisting of a type-I periodically-poled KTiOPO$_4$ (PPKTP) crystal inside an optical cavity and a laser pump may be used to realize the proposed heterodyne detection with $\omega_{\rm p}=\omega_{\rm s}+\omega_{\rm i}$, where $\omega_{\rm p}$ is the pump angular frequency. M$_1 \&$ M$_2$: Cavity mirrors. DM: dichromatic mirror.}
\label{fig:het}
\end{figure}

Inspired by the work on quantum noise cancellation of a parametric amplifier by correlating the amplifier's internal degree with the input signal through quantum entanglement \cite{kong2013}, we study heterodyne detection enhanced by quantum correlation between the image band vacuum and the signal, given that the image band vacuum may be thought of as the internal degree of a detector according to the theory of linear amplifier \cite{caves1982}. In this study, we focus on the following detection scenario: Prior to being sent to a heterodyne detector for detection, the signal light is firstly fed into a noiseless parametric amplifier \cite{ou1993}, whose pump light is at twice the frequency of the heterodyne local oscillator and hence quantum correlation is established between the signal mode and the image band (idler) mode in a vacuum state (figure \ref{fig:het}). Here we show that, if the amplified signal light is received by the heterodyne detector, the aforementioned 3 dB heterodyne noise can be completely eliminated due to quantum correlation. Moreover, our theoretical results show that optical amplification prior to signal detection may also serve to defeat the noise performance degradation in heterodyne detection due to imperfect quantum efficiency, akin to the case of direct detection \cite{bencheikh1995}.

Let consider a quantum field of signal light that has a continuum of frequency modes \cite{feng2016,glauber1963a,ou1987,mandel1995},
\begin{equation}\label{eq:field}
\hat{E}_{\rm s}^{(+)}(\mathbf{r},t)=\frac{\rm i}{\sqrt{\varepsilon_0V}} \sum_{\mathbf{k}}\left(\frac{1}{2}\hbar\omega_{\mathbf{k}}\right)^{\frac{1}{2}}
\hat{a}_{\mathbf{k}}{\rm e}^{{\rm i}(\mathbf{k}\cdot\mathbf{r}-\omega_{\mathbf{k}}t)},
\end{equation}
wherein $V$ stands for the quantization volume, $\varepsilon_0$ is the dielectric permittivity of vacuum, $\mathbf{k}$ is the set of plane-wave modes with $\omega_{\mathbf{k}}$ the corresponding angular frequency of each mode, and $\hbar\equiv h/2\pi$ in which $h$ represents the Planck constant. The amplitude operator $\hat{a}_{\mathbf{k}}$ is the photon annihilation operator for mode $\mathbf{k}$ and stays constant if there is no free electrical charge in the space \cite{glauber1963a}. The two mutually adjoint operators $\hat{a}_{\mathbf{k}}$ and $\hat{a}^\dagger_{\mathbf{k}}$ obey the following commutation relations,
\begin{equation}
\left[\hat{a}_{\mathbf{k}}, \hat{a}_{\mathbf{k'}}\right]=\left[\hat{a}^\dagger_{\mathbf{k}}, \hat{a}^\dagger_{\mathbf{k'}}\right] = 0,\
\left[\hat{a}_{\mathbf{k}}, \hat{a}^\dagger_{\mathbf{k'}}\right]=\delta_{\mathbf{k},\mathbf{k'}}.
\end{equation}

For simplicity, let further assume that the detected light is a single-frequency coherent field with an excited mode at the angular frequency of $\omega_{\rm s}$. If directly received by a detector with unity active detection area, the light field has an intrinsic signal-to-noise ratio (SNR),
\begin{equation}\label{eq:snrin}
\mbox{SNR}_{\rm in}=\frac{c\varepsilon_0}{\hbar \omega_{\rm s} B} <\hat{E}_{\rm s}^{(-)}(t)\hat{E}_{\rm s}^{(+)}(t)>
=\frac{c\varepsilon_0}{2\hbar \omega_{\rm s}B}|\alpha_{\rm s}|^2.
\end{equation}
Here $c$ stands for the speed of light in vacuum and $B$ represents the measurement bandwidth inversely proportional to the measurement time. $\alpha_{\rm s}\equiv<\hat{a}_{\rm s}> \times \sqrt{\hbar \omega_{\rm s}}/\sqrt{\varepsilon_0V}$ and $\hat{E}_{\rm s}^{(-)}(t)={\left[\hat{E}_{\rm s}^{(+)}(t)\right]}^\dagger$.

To quantitatively evaluate the noise performance of a heterodyne detector, we make use of the quantity of noise figure (NF),
\begin{eqnarray}\label{eq:nf}
\mbox{NF}=10\log_{10}\frac{\mbox{SNR}_{\rm in}}{\mbox{SNR}_{\rm out}},
\end{eqnarray}
where $\mbox{SNR}_{\rm out}$ is the signal SNR at the detector's output,
\begin{eqnarray}\label{eq:snrout}
\mbox{SNR}_{\rm out}\equiv\frac{P_{\rm out}}{\chi(\Omega)\cdot B}.
\end{eqnarray}
Here $P_{\rm out}$ is the average power of the output photoelectric signal produced by the detector and $\chi(\Omega)$ represents the average noise power density of the photoelectric signal at $\Omega=(\omega_{\rm s}-\omega_{\rm i}$)/2 ($\omega_{\rm i}$ stands for the angular frequency of the image band mode and $\omega_{\rm s}>\omega_{\rm i}$ is assumed; moreover, the frequency of the signal carried by the optical beam is much below the heterodyne frequency). A lower NF value indicates better noise performance for the detector.

From Eqs. (\ref{eq:snrin}), (\ref{eq:nf}) and (\ref{eq:snrout}), it follows that one needs the values of $P_{\rm out}$ and $\chi(\Omega)$ to calculate NF, the noise figure of the detector. The average signal power $P_{\rm out}$ may be figured out with
\begin{eqnarray}\label{eq:pave}
P_{\rm out}=\frac{1}{T}\int_0^T{\rm d} t J_-^2(t),
\end{eqnarray}
wherein $J_-(t)$ is the average differential photocurrent signal at the detector's output, \cite{feng2016,ou1987}
\begin{eqnarray}\label{eq:jminus}
J_-(t)=\eta \int^{\infty}_0 {\rm d} t' j(t')<\hat{I}_2(t-t')-\hat{I}_1(t-t')>,
\end{eqnarray}
to which a similar result may be obtained in semiclassical treatment \cite{mandel1995}. Here $\eta$ is the quantum efficiency of the detector in units of $(\hbar \omega_{\rm s})^{-1}$ (the average number of photoelectrons per photon energy), $j(t)$ is the output current pulse produced by a photoemission, and $j(t)=0$ for $t<0$. $\hat{I}_{1,2}(t)$ are the light intensities at the two output ports of the detector,
\begin{eqnarray}\label{eq:inthet}
\hat{I}_{1,2}(t)&=& (c\varepsilon_0/2)\{ \varepsilon_{\rm l}^{(-)}(t)\varepsilon_{\rm l}^{(+)}(t) + \hat{E}_{\rm a}^{(-)}(t)\hat{E}_{\rm a}^{(+)}(t)  \nonumber \\
&\pm& {\rm i}[\varepsilon_{\rm l}^{(+)}(t)\hat{E}_{\rm a}^{(-)}(t)-\varepsilon_{\rm l}^{(-)}(t)\hat{E}_{\rm a}^{(+)}(t)]\},
\end{eqnarray}
where $\varepsilon_{\rm l}^{(+)}(t)=\varepsilon_{\rm l}{\rm e}^{-{\rm i}\omega_{\rm l} t+{\rm i}\mathbf{k}_{\rm l}\cdot\mathbf{r}+{\rm i}\theta_{\rm l}}$ is the single-frequency classical field of the local oscillator (LO), with both the amplitude $\varepsilon_{\rm l}$ and phase $\theta_{\rm l}$ being real numbers. $\varepsilon_{\rm l}^{(-)}(t)={\left[\varepsilon_{\rm l}^{(+)}(t)\right]}^*$ and $\hat{E}_{\rm a}^{(+)}(t)$ is the field of the amplified signal produced in the amplifier (figure \ref{fig:het}b),
\begin{equation}\label{eq:afield}
\hat{E}_{\rm a}^{(+)}(\mathbf{r},t)=\frac{{\rm i}}{\sqrt{\varepsilon_0V}} \sum_{\mathbf{k}}\left(\frac{1}{2}\hbar\omega_{\mathbf{k}}\right)^{\frac{1}{2}}
\hat{b}_{\mathbf{k}}{\rm e}^{{\rm i}(\mathbf{k}\cdot\mathbf{r}-\omega_{\mathbf{k}}t)},
\end{equation}
which is related to the input light field $\hat{E}_{\rm s}^{(+)}(t)$ through the linear evolution equation \cite{caves1982},
\begin{equation}\label{eq:evolve}
\hat{b}_{\rm s}=\hat{a}_{\rm s}\cosh r + \hat{a}_{\rm i}^\dagger\sinh r,\hspace{0.1in}
\hat{b}_{\rm i}=\hat{a}_{\rm s}^\dagger\sinh r + \hat{a}_{\rm i}\cosh r.
\end{equation}
Here $\hat{b}_{\rm s,\rm i}$ is the photon annihilation operator of the signal (idler, or image band) mode of $\hat{E}_{\rm a}^{(+)}(\mathbf{r},t)$, and $r$ is a real constant determined by the strength and duration of the parametric amplification. From Eqs. (\ref{eq:field}), (\ref{eq:afield}), and (\ref{eq:evolve}), it is not difficult to show
\begin{eqnarray}\label{eq:fevolve}
\hat{E}_{\rm a}^{(+)}(\mathbf{r},t)&=&\hat{E}_{\rm s}^{(+)}(\mathbf{r},t)\cosh r\nonumber\\
&-&\hat{E}_{\rm i}^{(-)}(\mathbf{r},t){\rm e}^{2{\rm i}(\mathbf{k}_{\rm l}\cdot\mathbf{r}-\omega_{\rm l} t)}\sinh r ,
\end{eqnarray}
in which
\begin{equation}\label{eq:ifield}
\hat{E}_{\rm i}^{(+)}(\mathbf{r},t)\equiv\frac{\rm i}{\sqrt{\varepsilon_0V}}
\sum_{\mathbf{k}} \left(\frac{1}{2} \hbar |2\omega_{\rm l}-{\omega_{\mathbf{k}}}|\right)^{\frac{1}{2}}
\hat{a}_{\mathbf{k}}{\rm e}^{{\rm i}(\mathbf{k}\cdot\mathbf{r}-\omega_{\mathbf{k}}t)},
\end{equation}
provided that $\omega_{\rm s}+\omega_{\rm i}=2\omega_{\rm l}$ and $\mathbf{k}_{\rm s}+\mathbf{k}_{\rm i}=2\mathbf{k}_{\rm l}$.

Plugging Eqs. (\ref{eq:inthet}), (\ref{eq:fevolve}), and (\ref{eq:ifield}) into Eq. (\ref{eq:jminus}), one arrives at
\begin{eqnarray}\label{eq:jminusn}
&&J_-(t)=\sqrt{2}ce\varepsilon_0 \eta \varepsilon_{\rm l} |\alpha_{\rm s}| \nonumber\times\\
&&
\left[{\rm e}^r\cos\theta_{\rm l}\cos(\Omega t-\Delta \theta)-{\rm e}^{-r}\sin\theta_{\rm l}\sin(\Omega t-\Delta \theta)\right],\ \
\end{eqnarray}
where $\Delta \theta = \Delta\mathbf{k}\cdot\mathbf{r}-\theta_{\rm s}$, $\theta_{\rm s}$ is the phase of the signal mode, $\Delta \mathbf{k}\equiv \mathbf{k}_{\rm s}-\mathbf{k}_{\rm l}=\mathbf{k}_{\rm l}-\mathbf{k}_{\rm i}$, and we make use of $\omega_{\rm s}\approx \omega_{\rm i}$ since $|\omega_{\rm s}-\omega_{\rm i}|<<\omega_{\rm s,{\rm i}}$ for heterodyne detection. In addition, the detector assumes a sufficient response speed in photoemission, and hence $j(t)=e\delta(t)$ where $\delta(t)$ is Dirac function with $e$ being the charge on the electron.

From Eq. (\ref{eq:jminusn}), it follows that the photoelectric signal from the detector consists of a quadrature component $\sim \cos(\Omega t-\Delta \theta)$ that is amplified by a factor of ${\rm e}^r$ and a conjugate quadrature component $\sim \sin(\Omega t-\Delta \theta)$ that is reduced by the same factor, which holds true no matter what the input signal phase $\theta_{\rm s}$ is. When the LO phase $\theta_{\rm l}$ is controlled such that $\theta_{\rm l} = m\pi$ ($m$ any integer), the heterodyne detector produces an amplified signal $J_-(t)=\pm\sqrt{2}ce\varepsilon_0 \eta \varepsilon_{\rm l} |\alpha_{\rm s}| {\rm e}^r\cos(\Omega t-\Delta \theta)$ whose average power is, according to Eq. (\ref{eq:pave}),
\begin{eqnarray}\label{eq:pavec}
P_{\rm out}=(ce\varepsilon_0 \eta \varepsilon_{\rm l} |\alpha_{\rm s}|)^2 {\rm e}^{2r},
\end{eqnarray}
with an amplification gain of $G\equiv{\rm e}^{2r}$. On the other hand, if the LO phase $\theta_{\rm l} = (2m+1)\pi/2$, the heterodyne detector produces a reduced signal $J_-(t)=\pm\sqrt{2}ce\varepsilon_0 \eta \varepsilon_{\rm l} |\alpha_{\rm s}| {\rm e}^{-r}\sin(\Omega t-\Delta \theta)$. What is interesting is that the phase of the input signal, $\theta_{\rm s}$, does not need to be controlled, which is of technical essence for the studied detection scheme to be adapted to space-based GW experiments where violent signal phase disturbances are expected.

Next we proceed to calculate the noise power density $\chi(\Omega)$ of the heterodyne signal with the Fourier transform \cite{feng2016},
\begin{equation}\label{eq:chihet}
\chi(\omega)=\frac{1}{T}{\int}^{T}_{0}{\rm d} t{\int}^{+\infty}_{-\infty}{\rm d}\tau {\rm e}^{{\rm i}\omega \tau}<\Delta J_-(t) \Delta J_-(t+\tau)>,
\end{equation}
wherein the auto-correlation function of the differential-photocurrent fluctuations is \cite{ou1987}
\begin{eqnarray}\label{eq:autoJ}
&&<\Delta J_-(t) \ \Delta J_-(t+\tau)>\nonumber\\
&=&\sum_{\rm i=1}^2\eta \int^{\infty}_0 {\rm d} t'<\hat{I}_{\rm i}(t-t')>j_{\rm i}(t')j_{\rm i}(t'+\tau)\nonumber\\
&+&\sum_{\rm i,\rm j=1}^2 \eta^2(-1)^{\rm i+\rm j}\times \nonumber\\
&&\int\int_0^{\infty} {\rm d} t'{\rm d} t''j_{\rm i}(t')j_{\rm j}(t'') \lambda_{\rm i\rm j}(t-t',\tau+t'-t''),
\end{eqnarray}
which may be derived also under semi-classical approximation \cite{mandel1995}. Here $j_{\rm i,\rm j}(t)$ is photoemission-induced current pulse at the output ports of the detector, $\lambda_{\rm i\rm j}(t,\iota)\equiv <T:\Delta \hat{I}_{\rm i}(t)\Delta \hat{I}_{\rm j}(t+\iota):>$ are the correlation functions of light-intensity fluctuations, and the symbol $T: :$ means time- and normal-ordering of the field operators $\hat{E}_{\rm a}^{(\pm)}(t)$. Photodiode noise is not included in Eq. (\ref{eq:chihet}) since we consider only the situation in which the system sensitivity in the frequency band of interest is limited by the quantum noise of light \cite{Sesana2016}.

Under the approximations of a strong oscillator and fast response speed for the detector, the auto-correlation function (\ref{eq:autoJ}) may be readily reduced to
\begin{eqnarray}\label{eq:autoJJ}
&&<\Delta J_-(t) \ \Delta J_-(t+\tau)>\nonumber\\
&=&\eta c \varepsilon_0 e^2 \varepsilon_{\rm l}^2 \delta(\tau)+\eta^2e^2\sum_{\rm i,\rm j=1}^2 (-1)^{\rm i+\rm j} \lambda_{\rm i\rm j}(t,\tau).
\end{eqnarray}
Plugging Eq. (\ref{eq:autoJJ}) into Eq. (\ref{eq:chihet}) leads to
\begin{eqnarray}\label{eq:chi}
\chi(\omega)&=&\eta c \varepsilon_0 e^2 \varepsilon_{\rm l}^2+\eta^2e^2 \sum_{\rm i,\rm j=1}^2 (-1)^{\rm i+\rm j}\nonumber\\
&\times&\frac{1}{T}{\int}^{T}_{0}{\rm d} t{\int}^{+\infty}_{-\infty}{\rm d}\tau {\rm e}^{{\rm i}\omega \tau} \lambda_{\rm i\rm j}(t,\tau),
\end{eqnarray}
wherein the first term on the right hand side represents the detection shot noise.

From the definition of the correlation function $\lambda_{\rm i\rm j}(t,\tau)$ and Eq. (\ref{eq:inthet}), it follows that
\begin{eqnarray}\label{eq:app4}
&&\lambda_{\rm i\rm j}(t,\tau)=4^{-1}c^2 \varepsilon_0^2 (-1)^{\rm i+\rm j}\times\nonumber\\
&&\big[<\Delta\hat{E}_{\rm a}^{(-)}(t)\Delta\hat{E}_{\rm a}^{(+)}(t+\tau)>\varepsilon^{(+)}_{\rm l}(t)\varepsilon^{(-)}_{\rm l}(t+\tau) \nonumber \\
&+&<\Delta\hat{E}_{\rm a}^{(-)}(t+\tau)\Delta\hat{E}_{\rm a}^{(+)}(t)>\varepsilon^{(-)}_{\rm l}(t)\varepsilon^{(+)}_{\rm l}(t+\tau)\nonumber \\
&-&<\Delta\hat{E}_{\rm a}^{(-)}(t)\Delta\hat{E}_{\rm a}^{(-)}(t+\tau)>\varepsilon^{(+)}_{\rm l}(t)\varepsilon^{(+)}_{\rm l}(t+\tau) \nonumber \\
&-&<\Delta\hat{E}_{\rm a}^{(+)}(t+\tau)\Delta\hat{E}_{\rm a}^{(+)}(t)>\varepsilon^{(-)}_{\rm l}(t)\varepsilon^{(-)}_{\rm l}(t+\tau)\big] ,\nonumber\\
\end{eqnarray}
from which all the low-order terms in $\varepsilon_{\rm l}$ are dropped. With the help of the definitions of
\begin{eqnarray}\label{eq:gammad}
&&\Gamma_{\rm x}^{(1,1)}(t,\tau)\equiv <\Delta \hat{E}_{\rm x}^{(-)}(t)\Delta \hat{E}_{\rm x}^{(+)}(t+\tau)>{\rm e}^{{\rm i}\omega_{\rm l}\tau},\nonumber\\
&&\Gamma_{\rm x}^{(2,0)}(t,\tau)\equiv <\Delta \hat{E}_{\rm x}^{(-)}(t)\Delta \hat{E}_{\rm x}^{(-)}(t+\tau)>{\rm e}^{-{\rm i}\omega_{\rm l}(2t+\tau)},\nonumber\\
\end{eqnarray}
wherein ${\rm x=s,i,a}$, Eq. (\ref{eq:app4}) may be rewritten as
\begin{eqnarray}\label{eq:app5}
\lambda_{\rm i\rm j}(t,\tau)&=&4^{-1}c^2 \varepsilon_0^2 \varepsilon_{\rm l}^2 (-1)^{\rm i+\rm j}\nonumber\\
&\times& \big[\Gamma_{\rm a}^{(1,1)}(t,\tau)-\Gamma_{\rm a}^{(2,0)}(t,\tau){\rm e}^{2{\rm i}\theta'_{\rm l}}+ {\rm c.c.}\big],
\end{eqnarray}
in which $\theta'_{\rm l}\equiv\mathbf{k}_{\rm l}\cdot\mathbf{r}+\theta_{\rm l}$. From Eqs. (\ref{eq:chi}) and (\ref{eq:app5}), it follows that
\begin{eqnarray}\label{eq:chin}
\chi(\omega)&=&\eta c \varepsilon_0 e^2 \varepsilon_{\rm l}^2 +\eta^2 c^2 \varepsilon_0^2 e^2 \varepsilon_{\rm l}^2 \frac{1}{T}{\int}^{T}_{0}{\rm d} t{\int}^{+\infty}_{-\infty}{\rm d}\tau {\rm e}^{{\rm i}\omega \tau} \times\nonumber \\
&&
\big[\Gamma_{\rm a}^{(1,1)}(t,\tau)-\Gamma_{\rm a}^{(2,0)}(t,\tau){\rm e}^{2{\rm i}\theta'_{\rm l}}+ {\rm c.c.}\big].
\end{eqnarray}

In the following, we are going to evaluate $\Gamma_{\rm a}^{(1,1)}(t,\tau)$ and $\Gamma_{\rm a}^{(2,0)}(t,\tau)$ in Eq. (\ref{eq:chin}) using Eqs. (\ref{eq:fevolve}), (\ref{eq:ifield}), and (\ref{eq:gammad}). One may show without much difficulty that
\begin{eqnarray}\label{eq:gammaa}
\Gamma_{\rm a}^{(1,1)}(t,\tau)&=&\cosh^2r\ \Gamma_{\rm s}^{(1,1)}(t,\tau)\nonumber\\
&+&\sinh^2r \ {\rm e}^{-{\rm i}\omega_{\rm s}\tau} <\Delta \hat{E}_{\rm i}^{(+)}(t)\Delta \hat{E}_{\rm i}^{(-)}(t+\tau)>\nonumber\\
&-&\sinh r \cosh r \  \big[\Gamma^{(2,0)}(t,\tau){\rm e}^{2{\rm i}\mathbf{k}_{\rm l}\cdot\mathbf{r}}+{\rm c.c.}\big]
\nonumber\\
&=&\sinh^2r \ {\rm e}^{-{\rm i}\omega_{\rm l}\tau} <\big[\hat{E}_{\rm i}^{(+)}(t),\hat{E}_{\rm i}^{(-)}(t+\tau)\big]>,\nonumber\\
\end{eqnarray}
\begin{eqnarray}\label{eq:gammaaa}
\Gamma_{\rm a}^{(2,0)}(t,\tau)&=&\cosh^2r\ \Gamma_{\rm s}^{(2,0)}(t,\tau)\nonumber\\
&+&\sinh^2r\ {\rm e}^{-4{\rm i}\mathbf{k}_{\rm l}\cdot\mathbf{r}}\left[\Gamma_{\rm i}^{(2,0)}(t,\tau)\right]^*
\nonumber\\
&-&\sinh r \cosh r\times \nonumber\\
&&{\rm e}^{-{\rm i}(\omega_{\rm l}\tau+2\mathbf{k}_{\rm l}\cdot\mathbf{r})} <\Delta\hat{E}_{\rm i}^{(+)}(t)\Delta\hat{E}_{\rm s}^{(-)}(t+\tau)>\nonumber\\
&-&\sinh r\cosh r \times \nonumber\\
&&{\rm e}^{{\rm i}(\omega_{\rm l}\tau-2\mathbf{k}_{\rm l}\cdot\mathbf{r})}
<\Delta E_{\rm s}^{(-)}(t)\Delta E_{\rm i}^{(+)}(t+\tau)>\nonumber\\
&=&-\sinh r \cosh r\times \nonumber\\
&&{\rm e}^{-{\rm i}(\omega_{\rm l}\tau+2\mathbf{k}_{\rm l}\cdot\mathbf{r})} <\big[\hat{E}_{\rm i}^{(+)}(t),\hat{E}_{\rm s}^{(-)}(t+\tau)\big]>.\nonumber\\
\end{eqnarray}
Here
$\Gamma^{(2,0)}(t,\tau)\equiv <\Delta \hat{E}_{\rm s}^{(-)}(t)\Delta \hat{E}_{\rm i}^{(-)}(t+\tau)>{\rm e}^{-{\rm i}\omega_{\rm l}(2t+\tau)}$. In the last steps, $\Gamma_{\rm s}^{(1,1)}(t,\tau)=0$, $\Gamma^{(2,0)}(t,\tau)=0$,
$\Gamma_{\rm s,\rm i}^{(2,0)}(t,\tau)=0$,
$<\Delta \hat{E}_{\rm s}^{(-)}(t+\tau)\Delta \hat{E}_{\rm i}^{(+)}(t)>=0$, and \
$<\Delta \hat{E}_{\rm s}^{(-)}(t)\Delta \hat{E}_{\rm i}^{(+)}(t+\tau)>=0$
given that the fields $\hat{E}_{\rm s,\rm i}^{(+)}(t)$ are initially in coherent states \cite{glauber1963b}.

Although $\hat{E}_{\rm s,\rm i}^{(+)}(t)$ in Eqs. (\ref{eq:field}) and (\ref{eq:ifield}) are expressed in three dimensional (3D) expansions, all the above calculations hold valid for their one dimensional (1D) expansions as well. For optical fields in the form of collimated beams, one may substitute the 1D versions of Eqs. (\ref{eq:field}) and (\ref{eq:ifield}) into Eqs. (\ref{eq:gammaa}) and (\ref{eq:gammaaa}), leading to
\begin{eqnarray}\label{eq:gammab}
&&\Gamma_{\rm a}^{(1,1)}(t,\tau)-\Gamma_{\rm a}^{(2,0)}(t,\tau){\rm e}^{2{\rm i}\theta'_{\rm l}}\nonumber\\
&=&\frac{\hbar \sinh r}{2\pi c \varepsilon_0}
{\int}^{+\infty}_{0} {\rm d}\omega' {\rm e}^{{\rm i}(\omega'-\omega_{\rm l}) \tau}\times \nonumber\\
&&\left(\sinh r|2\omega_{\rm l}-\omega'|+{\rm e}^{2{\rm i}\theta_{\rm l}}\cosh r \sqrt{\omega'|2\omega_{\rm l}-\omega'|}\right),
\end{eqnarray}
after the summation over ${\rm k}$ is replaced by an integration: $(1/V) \sum_{\rm k} \rightarrow (1/2\pi)\int {\rm d} {\rm k}$ (${\rm k}=\pm\omega'/c$ is the wave number of light). Plugging Eq. (\ref{eq:gammab}) into Eq. (\ref{eq:chin}) and after some mathematical manipulations, one arrives at
\begin{eqnarray}\label{eq:chinn}
\chi(\omega)=\eta c \varepsilon_0 e^2 \varepsilon_{\rm l}^2\left[1+\eta\hbar\sinh r\
F(\omega)\right],
\end{eqnarray}
in which
\begin{eqnarray}\nonumber
F(\omega)&\equiv&\sinh r \ |\omega_{\rm l}+\omega|+\sinh r \ |\omega_{\rm l}-\omega| \nonumber\\ &&+\left({\rm e}^{2{\rm i}\theta_{\rm l}}\cosh r \sqrt{\omega_{\rm l}^2-\omega^2}+{\rm c.c.}\right)\ .
\end{eqnarray}

With higher amplification gains, stronger quantum correlations between the signal and image band (idler) modes are expected for better suppression of the 3 dB heterodyne noise. The gain may be limited by practically available LO power levels for the heterodyne detection, but a high gain of up to 45 dB is still allowed if a 20 mW LO is used for space-based GW searching \cite{luo2016}. Therefore, we will consider only the high-gain cases for NF calculations, i.e., $\sinh r\approx\cosh r\approx {\rm e}^r/2>>1$, and suppose that the LO phase is controlled for detection of the amplified (anti-squeezed) quadrature of the signal. In addition, the LO optical frequency is much higher than the heterodyne frequency, i.e., $\omega_{\rm l}>>\omega$. Under these approximations, Eq. (\ref{eq:chinn}) becomes
\begin{eqnarray}\label{eq:chinnn}
\chi(\omega)=2\eta c \varepsilon_0 e^2 \varepsilon_{\rm l}^2\left[1+(\eta\hbar\omega_{\rm l}){\rm e}^{2r}\cos^2\theta_{\rm l}\right].
\end{eqnarray}
The factor of 2 here accounts for the contribution of negative-frequency components when the calculation is compared with practical measurement \cite{feng2016}. From Eq. (\ref{eq:chinnn}) it follows that the heterodyne detector produces a maximal amplified signal at its output when $\theta_{\rm l}=0$, and the corresponding noise power level of the output signal is
\begin{eqnarray}\label{eq:chinnnn}
\chi(\omega)=2\eta c \varepsilon_0 e^2 \varepsilon_{\rm l}^2(\eta\hbar\omega_{\rm l}){\rm e}^{2r}.
\end{eqnarray}

From Eqs. (\ref{eq:snrout}), (\ref{eq:pavec}) and (\ref{eq:chinnnn}), it follows that the SNR of the amplified signal at the detector's output is
\begin{eqnarray}\label{eq:snroutn}
\mbox{SNR}_{\rm out}&=&\frac{P_{\rm out}}{\chi(\Omega)\cdot B}= \frac{(ce\varepsilon_0 \eta \varepsilon_{\rm l} |\alpha_{\rm s}|)^2 {\rm e}^{2r}}
{2\eta c \varepsilon_0 e^2 \varepsilon_{\rm l}^2(\eta\hbar\omega_{\rm l}){\rm e}^{2r} B}\nonumber\\
&=& \frac{c\varepsilon_0}{2\hbar\omega_{\rm l} B}|\alpha_{\rm s}|^2.
\end{eqnarray}
Using Eqs. (\ref{eq:snrin}), (\ref{eq:nf}) and (\ref{eq:snroutn}), one finally obtains the noise figure of the heterodyne detector,
\begin{eqnarray}\label{eq:nff}
\mbox{NF}=10\log_{10}\frac{\mbox{SNR}_{\rm in}}{\mbox{SNR}_{\rm out}}
=10\log_{10}\frac{\omega_{\rm l}}{\omega_{\rm s}}= 0 \ \mbox{dB},
\end{eqnarray}
where the approximation in the last step is based on the fact that $|\omega_{\rm l}-\omega_{\rm s}|/\omega_{\rm l}\rightarrow 0$ for heterodyne detection.

The result of Eq. (\ref{eq:nff}) proves that the noise performance of a heterodyne detector can be enhanced by the quantum correlation between the image band vacuum and the signal mode, without beating the quantum noise limit though. 
The price to pay though is the change of the phase sensitivity of the detector: The output signal becomes sensitive to the LO phase. The good news is that, no matter what the input phase $\theta_{\rm s}$ is, the amplifier will automatically amplify the $\cos(\Omega t-\Delta \theta)$ quadrature component of the detected signal with a gain of ${\rm G} = {\rm e}^{r}$, as shown by Eq. (\ref{eq:jminusn}). Therefore, the practical difficulty in the phase control for the input signal imposed by space-based GW experiments does not put any fundamental limit to the implementation of heterodyne detection enhanced by quantum correlation.

Another interesting feature in the studied heterodyne detection scheme revealed by Eq. (\ref{eq:nff}) is that the noise figure of the detector is independent of imperfect quantum efficiency $\eta$. It has been known for decades that the noise figure of a regular detector with imperfect quantum efficiency is \cite{bencheikh1995,mosset2005,pooser2009,corzo2012}
\begin{eqnarray}\label{eq:nfd}
\mbox{NF}=10\log_{10}\left(\xi^{-1}\right),
\end{eqnarray}
where $\xi=\eta\ \hbar\omega_{\rm s}<1$ in practice. In other words, a usual detector with lower $\eta$ entails higher NF and poorer noise performance. For direct detection scheme, optical amplification prior to signal detection may serve to conquer the noise figure degradation due to non-perfect quantum efficiency \cite{bencheikh1995}. Here in this work, we have shown the same effect for heterodyne detection, i.e., the ``beamsplitter noise" due to optical loss does not affect the SNR of a signal amplified by an amplifier without extra noises such as amplified spontaneous emission \cite{torounidis2006}. In the high-gain limit, Eq. (\ref{eq:nff}) shows that the noise figure of the heterodyne detector approaches its best value of 0 dB despite of imperfect quantum efficiency, from which space-based GW experiments will surely benefit.


This work is supported by the National Natural Science Foundation of China (grant No. 11947134 and 12074110).

\end{document}